\documentstyle[eqsecnum,preprint,aps,epsbox]{revtex}
\def\beq{\begin{equation}}
\def\eeq{\end{equation}}
\def\bea{\begin{eqnarray}}
\def\eea{\end{eqnarray}}
\def\nn{\nonumber\\}
\def\pa{\partial}
\def\cH{{\cal H}}
\def\cI{{\cal I}}
\def\cL{{\cal L}}
\def\cR{{\cal R}}
\def\cS{{\cal S}}
\def\mn{_{\mu\nu}}
\def\kt{{K^{\theta}_{\theta}}}
\def\bn{{\bar N}}
\def\br{{\bar R}}
\def\pt{\pa_{\tau}}
\def\pn{\pa_n}
\def\tr{\tilde R}
\def\tt{\tilde\tau}
\def\te{\tau_E}
\def\etal{{\it et al.\ }}
\def\ds{\displaystyle}

\begin{document}

\draft
\preprint{OCU-PHYS-241, AP-GR-31, gr-qc/0602084}
\title{The universe out of a monopole in the laboratory?
\footnote
{To appear in Phys. Rev. D with changing the title into "Is it 
possible to create a universe out of a monopole in the laboratory?"}}
\author{Nobuyuki Sakai$^{1,2,}$\thanks{Electronic address: nsakai@e.yamagata-u.ac.jp},
 Ken-ichi Nakao$^{3,}$\thanks{Electronic address: knakao@sci.osaka-cu.ac.jp},
Hideki Ishihara$^{3,}$\thanks{Electronic address: ishihara@sci.osaka-cu.ac.jp},
and Makoto Kobayashi$^{2}$}
\address{$^{1}$Department of Education, Yamagata University, Yamagata 990-8560, Japan}
\address{$^{2}$Department of Physics,   Yamagata University, Yamagata 990-8560, Japan}
\address{$^{3}$Department of Physics, Osaka City University, Osaka 558-8585, Japan}
\date{revised 30 June 2006}
\maketitle

\begin{abstract}
To explore the possibility that an inflationary universe can be created out of a 
stable particle in the laboratory, we consider the classical and quantum 
dynamics of a magnetic monopole in the thin-shell approximation.
Classically there are three types of solutions: stable, collapsing and 
inflating monopoles. We argue that the transition from a stable 
monopole to an inflating one could occur either by collision with a 
domain wall or by quantum tunneling.
\end{abstract}

\vskip 1cm
\begin{center}
PACS number(s): 04.60.Kz, 04.70.Bw, 14.80.Hv, 98.80.Cq
\end{center}

\newpage
\tighten

\section{Introduction}

For many years it has been discussed whether it is possible to create a universe
in the laboratory \cite{FG,FGG,FMP,GP99,GP01}.
The original idea was based on the model of a false-vacuum (de Sitter) bubble
embedded in an asymptotically flat (Schwarzschild) spacetime.
The classical dynamics of false-vacuum bubbles was originally 
studied by Sato, Sasaki, Kodama, and Maeda \cite{SSKM} in the context of 
old inflation \cite{SG}, and investigated systematically by Blau, 
Guendelman, and Guth \cite{BGG} with Israel's junction conditions 
\cite{Isr}.
If a false-vacuum bubble is larger than the de Sitter horizon, the 
bubble inflates eternally. Because the inflating bubble is surrounded 
by black hole horizons and causally disconnected by the ``original
universe", such a bubble is called a ``child universe".

Farhi and Guth \cite{FG} discussed whether such a false-vacuum bubble 
is created in the laboratory, applying 
the Penrose theorem \cite{Pen}. The theorem states that, if

(a) there exists a noncompact Cauchy surface,

(b) $\cR\mn k^{\mu}k^{\nu}\ge0$ for all null vector $k^{\mu}$,

(c) there exists an anti-trapped surface
\footnote{An anti-trapped surface is defined as the spacelike closed two-surface  
such that the expansion of both sets (i.e., ingoing and outgoing) of future 
directed null orthogonal to the two-surface is everywhere positive. This  
is just the past trapped surface introduced by Hayward \cite{Hay}.}
,\\
then there exists at least one past incomplete null geodesic. As a consequence of the 
Einstein equations $G\mn=8\pi GT\mn$, condition (b) is rewritten as
$T\mn k^{\mu}k^{\nu}\ge0$. Because any standard theory of matter, 
including a canonical scalar field, obeys this energy condition, we may 
conclude that it is impossible to create an inflationary universe in the laboratory.
Condition (c) represents the realization of an inflationary universe since 
the existence of an anti-trapped surface means the existence of the cosmological horizon. 
To put it simply, a false-vacuum bubble large enough to be an inflationary 
universe cannot avoid an initial singularity, while a bubble without 
an initial singularity is too small to expand.

Because the above argument is based on the classical field theory, 
a quantum process could make it possible to produce a large false-vacuum 
bubble without an initial singularity. Actually, Farhi, Guth, and Guven 
\cite{FGG} and Fischler, Morgan, and Polchinski \cite{FMP} considered 
a quantum decay from a small bubble without an initial singularity 
to a large bubble which becomes an inflationary universe, and 
calculated its probability.

As Guendelman and Portnoy \cite{GP99} pointed out, however, there is a 
problem in the model. Because the effective potential which governs 
the shell trajectories has no local minimum, there is no stable solution.
Even if we succeed to make a 
small false-vacuum bubble, the bubble collapses as soon as it is created;
there is almost no chance for a quantum decay to happen during its lifetime.
To solve this problem, Guendelman and Portnoy proposed a new model. They 
assumed a (2+1)-dimensional gauge field localized 
on the surface of a false-vacuum bubble. Due to the gauge field, there 
exists a static and stable classical configuration, which eventually 
decays into an inflationary universe.

Guendelman and Portnoy \cite{GP01} also proposed another model of a (2+1)-dimensional 
spacetime, where a massless scalar field localized at the 
(1+1)-dimensional boundary maintains a stable classical configuration.
A new aspect of this model is that an inflationary universe can be created 
by an arbitrarily small tunneling, which they called ``almost classical creation of a 
universe".

In this paper we consider the possibility that a stable magnetic 
monopole evolves into an inflationary universe.
In the Einstein-Yang-Mills-Higgs system static monopole solutions were 
intensively studied \cite{static}, and then dynamical solutions were also 
analyzed \cite{Sak}; one of the important results there is that there are 
stable solutions as well as inflating solutions with the same model 
parameters. We therefore expect the scenario that a classically stable monopole eventually 
evolves into an inflationary universe.

Although magnetic monopoles have never been detected, unified theories of elementary 
particles predict their existence. Furthermore, monopole inflation 
\cite{Sak,LVS,SNH,ACG}, 
which is free from the fine-tuning problem of initial conditions and 
the graceful exit problem, is still viable.
Therefore, the monopole model is more realistic and motivated than the previous models.

Specifically, we adopt the thin-shell model of Arreaga, Cho, and Guven 
\cite{ACG} (except for the form of the surface density).
A monopole is modeled as follows: the inside is de Sitter spacetime, the outside is 
Reissner-Nordstr\"om, and the boundary is a timelike hypersurface.
Here we should remark the limitation of this approximation. 
Numerical analysis of monopole inflation with the potential
$V=(\lambda/4)(\Phi^2-\eta^2)^2$ showed that the the boundary becomes 
spacelike once inflation begins \cite{SNH}. Nevertheless, the 
thin-shell model give a reliable result when a monopole oscillates 
stably or just begins to expand.
Because we are interested only in the transition from 
a stable state to an expanding state, the present model is effective.

The plan of this paper is as follows. In Sec. II we derive the 
classical action and the equation of motion of the shell.
In Secs. III and IV we consider the possibility that a classically stable 
monopole evolves into 
an inflationary universe by classical processes (Sec. III) and by 
quantum tunneling (Sec. IV). Sec. V is devoted to summary and 
discussions. In this paper we use the units $c=\hbar=1$, but 
occasionally we write $\hbar$ explicitly.

\section{Classical action and equations of motion}

The original model we consider is the SU(2) Einstein-Yang-Mills-Higgs system:
\beq\label{S}
\cS=\cS_g+\cS_m=\int d^4 x \sqrt{-g} \left[\frac{\cR}{16\pi G}
     -\frac14(F^a_{\mu\nu})^2-\frac12(D_{\mu}\Phi^a)^2-V(\Phi)\right],
\eeq
with
\begin{equation}\label{pote}
V(\Phi)= {1\over 4}\lambda(\Phi^2-\eta^2)^2, ~~ 
\Phi\equiv\sqrt{\Phi^a\Phi^a},
\end{equation} \begin{equation}
F^a\mn\equiv \partial_{\mu}A^a_{\nu}-\partial_{\nu}A^a_{\mu}
  +e\epsilon^{abc}A^b_{\mu}A^c_{\nu},~~
D_{\mu}\Phi^a\equiv\nabla_{\mu}\Phi^a+e\epsilon^{abc} 
A^b_{\mu}\Phi^c,
\end{equation}
where $A^a_{\mu}$ and $F^a_{\mu\nu}$ are the SU(2) Yang-Mills field 
potential and its field strength, respectively. $\Phi^a$ is the real
triplet Higgs field, and $V(\Phi)$ is its potential. $\lambda$ and 
$e$ are the Higgs self-coupling constant and the gauge coupling 
constant, respectively.
$\nabla_{\mu}$ and $D_{\mu}$ are the spacetime covariant derivative 
and the totally covariant derivative, respectively. 
The variation of (\ref{S}) with respect to $g_{\mu\nu}$, $\Phi^a$, and $A^a_i$ 
yields the Einstein equations
\beq\label{ein}
G\mn\equiv{\cal R}_{\mu\nu}-\frac12g_{\mu\nu}{\cal R}=8\pi GT\mn,
\eeq
\begin{equation}
T\mn\equiv D_{\mu}\Phi^aD_{\nu}\Phi^a
-g\mn\left[\frac12(D_{\sigma}\Phi^a)^2+V(\Phi)\right]
+F^a_{\mu\lambda}F^{a\lambda}_{\nu}-\frac14g_{\mu\nu}(F^a_{\lambda\sigma})^2,
\end{equation}
and the equations for the matter fields:
\begin{equation}\label{heq}
D_{\mu}D^{\mu} \Phi^a=\frac{\partial V(\Phi)}{\partial\Phi^a},
\end{equation}
\begin{equation}\label{geq}
D_{\mu}F^{a\mu\nu}=e\epsilon^{abc}\Phi^bD^{\nu}\Phi^c.
\end{equation}
We assume a spherically symmetric spacetime and adopt the 't Hooft-Polyakov 
ansatz for the matter field:
\begin{equation}\label{tp1}
\Phi^a=\Phi(x^0,x^1)\hat r^a,~~~
\hat r^a \equiv 
(\sin\theta\cos\varphi,\sin\theta\sin\varphi,\cos\theta),
\end{equation}
\begin{equation}\label{tp2}
A^a_{\mu}=\epsilon^{abc}(\pa_{\mu}\hat r^b)\hat r^c{1-w(x^0,x^1)\over e}.
\end{equation}

The purpose of this section is to reduce the action (\ref{S}) to its 
thin-shell limit and derive the equations of motion.
We essentially follow Farhi \etal \cite{FGG} and Ansoldi \etal \cite{AABS} 
except for treatment of the boundary of the region of integration.
We will not introduce the boundary term 
to cancel out the second derivatives of the metric; instead we will delete the second
derivatives just by integration by part with fixing all dynamical variables 
and their first derivatives at the boundary. Because the two methods are 
equivalent, we choose this simple and straightforward way.

In the thin-shell limit, the outside ($V_+$) and the inside ($V_-$) 
are characterized by $\Phi=\eta,~w=0$ and $\Phi=0,~w=1$, respectively. 
Then, the spacetime solutions and the energy-momentum tensors of both sides 
are respectively given by
\beq\label{RN}
ds^2=-A_+dt_+^2+{dr_+^2\over A_+}+r_+^2(d\theta_+^2+\sin^2\theta_+d\varphi_+^2),
~~~ A_+(r_+)\equiv1-{2GM\over r_+}+{GQ^2\over r_+^2}.
\eeq\beq
T_{\nu}^{\mu+}={1\over 2e^2r_+^4}{\rm diag}(-1,-1,1,1),~~~
Q^2={4\pi\over e^2}
\eeq
\beq\label{DS}
ds^2=-A_-dt_-^2+{dr_-^2\over A_-}+r_-^2(d\theta_-^2+\sin^2\theta_-d\varphi_-^2),
~~~ A_-(r_-)\equiv1-H^2r_-^2,
\eeq\beq
T_{\nu}^{\mu-}=-\rho\delta^{\mu}_{\nu},~~~H^2={8\pi G\over 3}\rho,
\eeq
where $\rho=V(0)=\lambda\eta^4/4$ is a constant.
The two regions are connected at the $r_-=r_+=R$ spherical hypersurface 
$\Sigma$. Because we can identify $(r_-,\theta_-,\varphi_-)$ and 
$(r_+,\theta_+,\varphi_+)$ on $\Sigma$, hereafter we omit the sings $\pm$ 
in these coordinates.

The matter part (of the action) in $V^{\pm}$ is evaluated as
\bea\label{Smp}
\cS_m^+&=&\int^{t^+_f}_{t^+_i}dt_+\int^{\infty}_Rdr~4\pi r^2
\left(-{1\over 2e^2r^4}\right)
=-\int^{t^+_f}_{t^+_i}dt_+~{Q^2\over2R},\\
\label{Smm}
\cS_m^-&=&\int^{t^-_f}_{t^-_i}dt_-\int^{R}_0 dr~4\pi r^2(-\rho)
=-\int^{t^-_f}_{t^-_i}dt_-{4\pi\rho\over 3}R^3,
\eea
where $[t^{\pm}_i,t^{\pm}_f]$ is the time interval under consideration.
For the gravity part in $V^{\pm}$, we apply the Einstein equation 
(\ref{ein}), which reads $\cR^{\pm}=-8\pi GT^{\pm}$. Then we find
\beq\label{Sg}
\cS_g^+=0,~~~
\cS_g^-=\int^{t^-_f}_{t^-_i}dt_-\int^{R}_0 dr~4\pi r^2
{-8\pi G\over 16\pi G}(-4\rho)
=\int^{t^-_f}_{t^-_i}dt_-~{8\pi\rho\over 3}R^3.
\eeq

To describe the geometry in the neighborhood of $\Sigma$ we introduce the 
Gaussian normal coordinate system:
\beq\label{GN}
ds^2=dn^2+\gamma_{ij}dx^idx^j
=-\bn(\tau,n)^2d\tau^2+dn^2+\br(\tau,n)^2(d\theta^2+\sin^2\theta d\varphi^2),
\eeq
and define the metric functions on the shell ($\Sigma: n=0$) as
\beq
N(\tau)\equiv\bn(\tau,0),~~~ R(\tau)\equiv\br(\tau,0).
\eeq
Following Ansoldi \etal \cite{AABS}, we keep $N$ an arbitrary function
to derive the constraint equation from the variational principle.

The matter part on $\Sigma$ should be derived from the original action 
(\ref{S}). We depart from Arreaga \etal \cite{ACG} by 
considering the Yang-Mills gauge field as well.
Because the dominant contribution on the shell comes from gradient 
energy, the matter part of the action approximates
\beq
\cS_m\approx\int^{\tau_f}_{\tau_i}d\tau\int^{+0}_{-0}dn~
4\pi\bn\br^2\left\{-{(\pa_n\Phi)^2\over2}-\left({\pa_nw\over e\br}\right)^2\right\}.
\eeq
Defining two functions of $\tau$ (definite integrals of $n$) as
\beq
\sigma_0\equiv\int^{+0}_{-0}dn~{(\pa_n\Phi)^2\over2},~~~
\sigma_1\equiv\int^{+0}_{-0}dn~{(\pa_nw)^2\over e^2},
\eeq
the action in the thin-shell limit becomes
\beq\label{Sms}
\cS_m^{\Sigma}=\int^{\tau_f}_{\tau_i}N d\tau(-4\pi\sigma R^2),~~~
\sigma(\tau)\equiv\sigma_0(\tau)+{\sigma_1(\tau)\over R^2(\tau)}.
\eeq

To evaluate the gravity part for $\Sigma$, we define the extrinsic curvature tensor as
\beq
K_{ij}\equiv\nabla_jn_i=-\Gamma^n_{ij}={\pa_n\gamma_{ij}\over 2}
~~~ {\rm at} ~~~ n\rightarrow\pm0,
\eeq
where $n^{\mu}$ is the normal vector of $\Sigma$ pointing outward, and
given by $n^n=1$ and $n^i=0$ in the coordinate system (\ref{GN}).
Then we can decompose the four dimensional Ricci scalar 
into the three dimensional Ricci scalar and the extrinsic curvature:
\bea\label{Sgs}
\cS_g^{\Sigma}
&=&\int^{\tau_f}_{\tau_i}\bn d\tau\int^{+0}_{-0} dn~4\pi\br^2{1\over 16\pi G}
(\cR^{(3)}-K_{ij}K^{ij}-K^2-2\pa_nK) \nn
&=&\int^{\tau_f}_{\tau_i}N d\tau\left(-{R^2\over 2G}[K]^{\pm}\right),
\eea
where $K\equiv K^i_i$ and $[K]^{\pm}\equiv K^+-K^-$. The components 
of $K_{ij}$ for the line elements (\ref{DS}) and (\ref{RN}) are calculated as
\bea
\kt^{\pm}&=&{\beta^{\pm}\over R},~~~~~
\beta^{\pm}\equiv\pa_n\br^{\pm}=\varepsilon^{\pm}\sqrt{\dot R^2+A_{\pm}(R)},\\
{K^{\tau}_{\tau}}^{\pm}&=&{1\over\beta^{\pm}}\left(\ddot R+{A_{\pm}'(R)\over2}\right),
\eea
where 
\beq
\varepsilon^{\pm}=+1 ~ {\rm or} ~ -1,
~~~\dot{~}\equiv\frac1N{d\over d\tau}, ~~~ '\equiv{d\over dr}.
\eeq

From {(\ref{Smp})-{(\ref{Sg}), {(\ref{Sms}) and {(\ref{Sgs}), 
the total action is reduced to
\beq\label{S2}
\cS=-\int^{t^+_f}_{t^+_i}dt_+~{Q^2\over 2R}
+\int^{t^-_f}_{t^-_i}dt_-~{4\pi\rho\over 3}R^3
-\int^{\tau_f}_{\tau_i}Nd\tau\left\{4\pi\sigma R^2
+\frac1G\left[{R^2\over2\beta}\left(\ddot R+{A'\over2}\right)+R\beta\right]^{\pm}\right\}
\eeq
To remove the second-derivative term from the action, we integrate it by 
part (with respect to the proper time $\ds T\equiv\int Nd\tau$):
\beq
\int^{\tau_f}_{\tau_i}Nd\tau{R^2\ddot R\over2\beta}
={R^2\over2}\tanh^{-1}\left({\dot R\over\beta}\right)\Bigg|^{T_f}_{T_i}
+\int^{\tau_f}_{\tau_i}Nd\tau\left\{
-R\dot R\tanh^{-1}\left({\dot R\over\beta}\right)
+{R^2\dot R^2A'\over4\beta A}\right\}
\eeq
Because the first term in the RHS contains only variables at the 
boundary, we can ignore it in variation.
We also note the relation,
\beq
\int^{\tau_f}_{\tau_i}N d\tau{R^2A'_+\beta_+\over 4A_+}
=\int^{t^+_f}_{t^+_i}dt_+{R^2A'_+\over 4}
={GM\over 2}(t^+_f-t^+_i)-\int^{t^+_f}_{t^+_i}dt_+{GQ^2\over 2R}.
\eeq
Again we can ignore the first term in the RHS in variation, and the 
second term there is canceled by the first term in (\ref{S2}). Thus we 
arrive at the final form of the action:
\beq\label{S3}
\cS=\int^{\tau_f}_{\tau_i}d\tau\cL,~~~
\cL\equiv -N\left\{4\pi\sigma R^2
+\frac RG\left[\beta-\dot R\tanh^{-1}\left({\dot R\over\beta}\right)\right]^{\pm}\right\}.
\eeq

Keeping in mind that $N$ is hidden in $\dot R\equiv(dR/d\tau)/N$,
the variation of (\ref{S3}) with respect to $N$ and $R$ yields the 
classical equations of motion:
\beq\label{jc1}
[\kt]^{\pm}\equiv{[\beta]^{\pm}\over R}
=-4\pi G\left(\sigma_0+{\sigma_1\over R^2}\right),
\eeq\beq\label{jc2}
[K^{\tau}_{\tau}]^{\pm}\equiv{[\dot\beta]^{\pm}\over\dot R}
=-4\pi G\left(\sigma_0-{\sigma_1\over R^2}\right).
\eeq
These equations give the energy-momentum conservation on the shell,
\beq\label{emc}
\dot\sigma_0+{\dot\sigma_1\over R^2}=0.
\eeq
The original field equations (\ref{heq}) and (\ref{geq}), however, give
further conditions,
\beq\label{const}
\dot\sigma_0=\dot\sigma_1=0,
\eeq
as is shown in Appendix.

Now we define the conjugate momentum as
\beq\label{P}
P\equiv{\pa\cL\over\pa\dot R}=\frac{NR}G
\left[\tanh^{-1}\left({\dot R\over\beta}\right)\right]^{\pm},
\eeq
and the Hamiltonian as
\beq\label{H}
\cH\equiv P\dot R-\cL=\frac{NR}G(\beta_+-\beta_-+4\pi G\sigma R).
\eeq
The equations of motion (\ref{jc1}) and (\ref{jc2}) are equivalent to 
the Hamiltonian constraint:
\beq\label{HC}
\cH=0~~~{\rm with}~~~\sigma_0,~\sigma_1={\rm const.}
\eeq
Formally one should invert (\ref{P}) to express $\dot R$ in terms of 
$P$ in order to eliminate $\dot R$ in (\ref{H}), as Ansoldi \etal did \cite{AABS}.
In Sec. IV, however, we will quantize the system without the explicit form 
of $\cH(P,R)$. 

\section{Classical dynamics}

Hereafter we take $N=1$.
The classical motion is governed by the constraint equation (\ref{HC}), 
or (\ref{jc1}).
Following Arreaga \etal \cite{ACG}, we introduce dimensionless quantities, 
\beq\label{rescale}
\tr\equiv HR,~~~ \tt\equiv H\tau,~~~
m\equiv HGM,~~~ q^2\equiv H^2GQ^2,~~~
s_0\equiv{4\pi G\sigma_0\over H},~~~
s_1\equiv4\pi GH\sigma_1,
\eeq
to rewrite Eq.(\ref{jc1}) as
\beq\label{eom}
\left({d\tr\over d\tt}\right)^2+U(\tr)=-1,
\eeq\beq\label{V}
U(\tr)\equiv-\left({1-s_0^2\over2}\tr-{s_0s_1\over\tr}-{m\over\tr^2}
+{q^2-s_1^2\over2\tr^3}\right)^2\left(s_0+{s_1\over\tr^2}\right)^{-2}-\tr^2.
\eeq

To understand the global spacetime structure, it is helpful to know 
the positions of horizons and the signs of $\beta^{\pm}=\pa_nr_{\pm}$ in 
terms of $\tr$.  De Sitter horizons $\tr_D$, the black-hole outer 
horizons $\tr_{(+)}$ and the inner horizons $\tr_{(-)}$ are characterized by
\beq
\tr_D=1,~~~ \tr_{(\pm)}=m\pm\sqrt{m^2-q^2},
\eeq
respectively. To clarify the sings of $\beta^{\pm}$, using 
(\ref{jc1}), we reexpress them as
\beq\label{beta}
\beta^{\pm}=\left(\frac{\mp s_0^2-1}{2}\tilde{R}\mp\frac{s_0 s_1}{\tilde{R}}
+\frac{m}{\tilde{R}^2}-\frac{q^2\pm s_1^2}{2\tilde{R}^3}\right)
\left(s_0+\frac{s_1}{\tilde{R}^2}\right)^{-1}.
\eeq

The conformal diagrams for de Sitter spacetime and Reissner-Nordstr\"om 
spacetime are shown in Fig.\ 1. We shall describe monopole solutions 
by joining a part of de Sitter spacetime to a part of 
Reissner-Nordstr\"om. The boundary of the two parts represents the 
trajectory of the shell $\Sigma$. 
We choose the monopole center as the ``left" $r=0$ line in the 
diagram (a). Accordingly, the normal vector $n^{\mu}$ of $\Sigma$ 
points to the right in all diagrams. 
In the region $r_-<r_D$ ($r_+>r_{(+)}$) the sign of 
$\beta^{-}=\pa_nr_{-}$ ($\beta^{+}=\pa_nr_{+}$) is definite
regardless of the shell motion.
Inversely, the signs of (\ref{beta}) 
tell us approximate trajectories of the shell without solving the 
equation of motion (\ref{eom}).

There are four parameters, $m,~q,~s_0$, and $s_1$, which 
should be determined by the model parameters in (\ref{S}) and initial 
conditions. From the field equations derived from (\ref{S}), we can 
estimate their order-of-magnitude as
\beq
q^2\sim{\lambda\over e^2}\left({\eta\over m_{Pl}}\right)^4,~~~
s_0\sim{\eta\over m_{Pl}},~~~
s_1\sim{\sqrt{\lambda}\over e}\left({\eta\over m_{Pl}}\right)^3,
\eeq
where $m_{Pl}\equiv1/\sqrt{G}$ is the Planck mass.
For example, if we assume $\eta/m_{Pl}\sim1$ and $\lambda/e^2\sim1$, 
all of them should be of order of unity.
The precise values of those parameters cannot be determined
without solving the field equations by fixing $\Phi$ and $\pa_t\Phi$ at $t=0$; 
there remains some ambiguity.
We therefore regard the four parameters as free parameters in the range 
of order of unity.
Here we do not survey all classical solutions, but 
only show some of them and discuss whether stable monopoles can evolve 
into inflating monopoles without an initial singularity, where we mean ``initial 
singularity" by the spacetime singularity which exists in the past of the experimenter 
who makes an inflating monopole.

Figure 2 shows a classically stable oscillating monopole in a horizonless spacetime 
(type A) with $m=0.58,~q=0.6,~s_0=0.6$ and $s_1=0.1$. In the same potential there 
is another expanding solution for large $\tr$ (type A'). Static and 
stable solutions obtained without thin-shell approximation \cite{static} 
correspond to Type A.

If we increase $m$, the feature of solutions changes drastically. 
Figure 3 shows an inflating monopole (type B) with $m=0.64,~q=0.6,~s_0=0.6$ and $s_1=0.1$.
This phenomenon is consistent with the previous result \cite{static,Sak} 
that static monopole solutions are nonexistent if their gravitational 
mass are large enough.

We expect that type A monopoles can evolve into type B by
accretion of mass to the monopole.
Specifically, we consider the model that a spherical domain wall 
surrounding the monopole eventually collides with it. Possible trajectories 
before and after the collision is shown in Fig.\ 4. This could be a classical 
process that an inflationary universe is created in the laboratory.

What about an initial singularity?
In agreement with Farhi and Guth, this created universe includes 
past incomplete null geodesics emanating from anti-trapped surfaces, 
as is shown in Fig.\ 4. As one can easily see from 
Fig.\ 4, however, there is no initial singularity such as the Big Bang.
Although a singularity exists in the past of the inflating monopole, 
the singularity is located in the future of the experimenter in the laboratory.
In other words, even if no singularity exists in the past of the experimenter who 
makes a monopole, inflation in the monopole is realizable in the future of the experimenter. 
From a observational point of view, however, since the inflating monopole 
is realized inside a black hole, the experimenter cannot 
observe it unless he or she enters into the black hole. 
The detectability and stability of this solution will be discussed in the final section. 

In the intermediate case between A and C solutions, there are 
classically stable oscillating solutions with black-hole horizons. 
Figure 5 shows the case of $m=0.61,~q=0.6,~s_0=0.6$ and $s_1=0.1$.
In this case there are two types of classical solutions: a stable 
oscillating monopole (type C) and an inflating monopole (type C').
Type C solutions are stable but do not fall into any solution in the 
study of static solutions \cite{static}. It was found that stably oscillating 
solutions exist in the parameter range where static solutions are 
nonexistent \cite{Sak}.
Quantum tunneling from C to C' is the subject of the next section.

\section{Quantum tunneling}

To quantize the system we define the operators as
\beq
\hat R\equiv R,~~~ \hat P\equiv i\hbar{\pa\over\pa R},~~~
\hat\cH\equiv\cH(\hat P,\hat R).
\eeq
and impose the Hamiltonian constraint on the quantum state $\Psi$,
\beq\label{WD}
\hat\cH(\hat P,\hat R)\Psi(R)=0.
\eeq
If we write the wave function as
\beq
\Psi(R)=e^{iF(R)/\hbar},
\eeq
and substitute into (\ref{WD}), to lowest order in the WKB expansion, 
we obtain the Hamilton-Jacobi equation:
\beq
\cH\left({dF\over dR},R\right)=0.
\eeq
The solution $F(R)$ is given by
\beq
F(R)=\int^RP(R)dR=\int^{\tau}d\tau R\dot R
\left[\tanh^{-1}\left({\dot R\over\beta}\right)\right]^{\pm},
\eeq
which is nothing but the action (\ref{S3}) with the Hamiltonian constraint 
$\cH=0$.

In the classically forbidden region we assume that there is a 
solution $R(\te)$ to the classical equation of motion, where $\te$ is 
the Euclidean time defined as
\beq
\te\equiv i\tau.
\eeq
We also define the Euclidean action as
\beq
F_E(R(\te))\equiv iF(R(i\tau))=
\int^{\te}d\te R{dR\over d\te}\left[\tan^{-1}\left({dR/d\te\over\beta_E}\right)\right]^{\pm},
\eeq
where
\beq
\beta_E\equiv\beta\left(\tau\rightarrow{\te\over i}\right)
=\varepsilon\sqrt{-\left({dR\over d\te}\right)^2+A}.
\eeq
Then the ratio of amplitudes at $R_i$ and $R_f$ is given by
\beq\label{ratio}
{\Psi(R_f)\over\Psi(R_i)}\approx\exp\left(-{B\over\hbar}\right),
\eeq\beq\label{B}
B\equiv F_E(R_f)-F_E(R_i)=\frac1G
\int^{\te^f}_{\te^i}d\te R{dR\over d\te}
\left[\tan^{-1}\left({dR/d\te\over\beta_E}\right)\right]^{\pm}.
\eeq

To integrate (\ref{B}) we need to solve the Euclidean equation of 
motion, which is the analytic continuation of (\ref{jc1}):
\beq\label{jcE}
\beta_E^+-\beta_E^-=-4\pi G\sigma R,
\eeq
Rescaling the quantities as (\ref{rescale}) and $\tt_E\equiv H\tau$, we can rewrite 
(\ref{jcE}) as
\beq\label{eomE}
\left({d\tr\over d\tt_E}\right)^2=U(\tr)+1.
\eeq
In numerical calculation, it is more convenient to integrate its derivative,
\beq\label{eomE2}
{d^2\tr\over d\tt_E^2}=\frac12{dU\over d\tr},
\eeq
once initial values are given by (\ref{eomE}). 
The Euclidean junction condition (\ref{jcE}) also gives
\beq\label{betaE}
\beta_E^{\pm}=\left(\frac{\mp s_0^2-1}{2}\tilde{R}\mp\frac{s_0 s_1}{\tilde{R}}
+\frac{m}{\tilde{R}^2}-\frac{q^2\pm s_1^2}{2\tilde{R}^3}\right)
\left(s_0+\frac{s_1}{\tilde{R}^2}\right)^{-1}.
\eeq
which is identical to the expression (\ref{beta}).
The coefficient $B$ is also rewritten with the normalized variables as 
\beq\label{BB}
B=\frac1{GH^2}\int^{\tt_E^f}_{\tt_E^i}d\tt_E\tr{d\tr\over d\tt_E}
\left[\tan^{-1}\left({d\tr/d\tt_E\over\beta_E}\right)\right]^{\pm},
\eeq

The method of numerical calculation of $B$ is as follows. (i) Give
the initial values $\tr$ and $d\tr/d\tt_E=0$ (at $\tt=\tt_E^i$), which 
satisfies (\ref{eomE}). (ii) Integrate (\ref{eomE2}) and (\ref{BB}) 
with (\ref{betaE}) until $d\tr/d\tt_E=0$ again (at $\tt=\tt_E^f$).
(iii) Check the accuracy of numerical solutions with (\ref{eomE}).

Now let us investigate quantum tunneling from a classically stable monopole 
to inflating one. The first candidate is a quantum decay from type A to 
A'; however, it turns out to be impossible.
Because $\beta_+$ is always negative in type A', a possible spacetime 
structure is given by Fig.\ 2(c); the direction of the outer Reissner-Nordstr\"om 
region is opposite to that of type A in Fig.\ 2(b). It seems 
impossible that the outer infinite region becomes finite with a naked singularity by quantum 
tunneling. Actually, because a part of the action, given by (\ref{S}), diverges 
if the outer region contains $r=0$, $B$ also diverges and the 
probability $\sim e^{-B}$ becomes zero.

On the other hand, the tunneling from type C to C' could happen. 
Therefore, we calculate $B$ for this decay. Figure 6 shows some of the 
results, which are normalized by $1/GH^2$. The normalization factor is 
given by $1/GH^2=(3/2\pi\lambda)(m_{Pl}/\eta)^4$ in the model (\ref{S}).
Therefore, if $\lambda\eta^4$ is not much smaller than the Planck density, $B$ is 
not much larger than unity. In this case, the probability $e^{-B}$ is 
considerable and the tunneling from type C to C' is likely to happen in the
laboratory.

\section{Summary and Discussions}

To explore the possibility that an inflationary universe can be created out of a 
stable particle in the laboratory, we have considered the classical and quantum 
dynamics of a magnetic monopole in the thin-shell approximation.

There are two advantages in the monopole model, compared with the 
false-vacuum model. First, magnetic monopoles are natural 
consequences of particle theories; monopole inflation is also one of the
consequences and still viable in cosmology.
Second, contrary to the model of a false-vacuum bubble, there are stable 
solutions besides inflating and collapsing solutions in the present
model. This is a preferable feature for making a universe in real experiments.

It has been believed that, as Farhi and Guth argued, the Penrose theorem 
indicates impossibility of creation of an inflationary universe without an initial 
singularity, and hence quantum tunneling has been devised to escape 
from the theorem. 
We have found, however, that in agreement with Farhi 
and Guth, there are past incomplete null geodesics, 
but the inflating monopole could be created by the 
experimenter whose past is geodesically complete.
We have proposed a specific model that a domain wall surrounding the 
stable monopole coalesces with it and becomes an inflating monopole.

For the trajectories beyond the inner horizon, or the Cauchy horizon, 
we should discuss instability of the horizon. Poisson and 
Israel \cite{PI} argued that, if radial perturbations are given, the gravitational 
mass inside the Cauchy horizon increases infinitely, which leads the 
appearance of a spacelike singularity near the Cauchy horizon. If 
perturbations are given in the present model for some physical reasons,
classical solutions beyond the Cauchy horizon may break down. According to Dafermos \cite{Daf}, 
however, the spacetime could be extendible beyond such singularity as 
a $C^0$ metric. Therefore, a physical consequence of our classical 
solutions is still unclear.

We have also analyzed a quantum decay from a classically stable monopole to an 
inflating one. We have adopted the canonical quantization of Farhi 
\etal to evaluate the probability amplitude to lowest order in WKB 
approximation. We find that, if the energy scale of the model is close 
to the Planck scale, the probability amplitude is considerable and the 
tunneling from type C to C' is likely to happen in the laboratory.

A problem of this model and other related models is the difficulty
of detecting an inflationary universe because it is surrounded by an 
event horizon, which eventually disappears by Hawking radiation.
Recently, Hawking \cite{Haw} argued that information is preserved in 
black hole formation and evaporation, and information could get out 
of a black hole by radiation. Although this conjecture is uncertain at the 
moment, we expect that it will be a clue to detect a universe in the laboratory.

\acknowledgements
We thank M. Dafermos, H. Kodama, R. Myers, and M. Sasaki for useful discussions.
The substantial part of this work was done while NS stayed at Department of Physics, 
Osaka City University. NS thanks his colleagues there for their hospitality.
This work was supported in part by MEXT KAKENHI Nos.\ 14540275, 15740132, 
16540264 and 18540248.

\appendix
\section{Proof of (2.34)}

The field equations (\ref{heq}) and (\ref{geq}) with the Gaussian normal 
coordinates (\ref{GN}) are written as
\beq\label{emcp1}
 {1\over\bn\br^2}\pt\left({\br^2\over\bn}\pt\Phi\right)
-{1\over\bn\br^2}\pn(\bn\br^2\pn\Phi)
+{2\over\br^2}w^2\Phi+{dV\over d\Phi}=0,
\eeq\beq\label{emcw1}
 {1\over\bn}\pt\left({\pt w\over\bn}\right)
-{1\over\bn}\pn(\bn\pn w)+e^2\Phi^2w+{w(w^2-1)\over\br^2}=0.
\eeq
Multiplying (\ref{emcp1}) by $\pt\Phi$ and (\ref{emcw1}) by $\pt w$, it follows that
\beq\label{emcp2}
{1\over\br^4}\pt\left\{\frac12\left({\br^2\over\bn}\pt\Phi\right)^2\right\}
+\pt\left\{{(\pn\Phi)^2\over2}\right\}
-{\pn(\bn\br^2\pt\Phi\pn\Phi)\over\bn\br^2}
+\pt\Phi\left({2\over\br^2}w^2\Phi+{dV\over d\Phi}\right)=0,
\eeq\beq\label{emcw2}
\pt\left\{\frac12\left({\pt w\over\bn}\right)^2+{(\pn w)^2\over2}\right\}
-{\pn(\bn\pt w\pn w)\over\bn}
+w\pt w\left(e^2\Phi^2+{w^2-1\over\br^2}\right)=0.
\eeq

In the thin-shell limit we have assumed
\beq
{(\pn\Phi)^2\over2}=\sigma_0(\tau)\delta(n),~~~
(\pn w)^2=e^2\sigma_1(\tau)\delta(n),
\eeq
while $\Phi,~\pt\Phi,~w$ and $\pt w$ are finite on the shell. Therefore, the 
integration of (\ref{emcp2}) and (\ref{emcw2}) with $\ds\int^{+0}_{-0}dn$ yields
\beq\label{emcp3}
{d\sigma_0\over d\tau}-[\pt\Phi\pn\Phi]^{\pm}
+\int^{+0}_{-0}dn{\pn(\bn\br^2)\over\bn\br^2}\pt\Phi\pn\Phi=0,
\eeq\beq\label{emcw3}
{e^2\over2}{d\sigma_1\over d\tau}-[\pt w\pn w]^{\pm}
+\int^{+0}_{-0}dn{\pn\bn\over\bn}\pt w\pn w=0.
\eeq
Although $\pn\Phi$ and $\pn w$ diverge at $n=0$, their linear terms 
make no contribution to the integral because of
$\ds\int^{+0}_{-0}dn\sqrt{\delta(n)}=0$.
Hence, the third terms in (\ref{emcp3}) and in (\ref{emcw3}) vanish.
Furthermore, because $\Phi^{\pm}$ and $w^{\pm}$ are constant in the present 
model, the second terms in (\ref{emcp3}) and in (\ref{emcw3}) also vanish.
Thus (\ref{const}) has been proved.



\begin{figure}
 \begin{center}
  \psbox[scale=0.7]{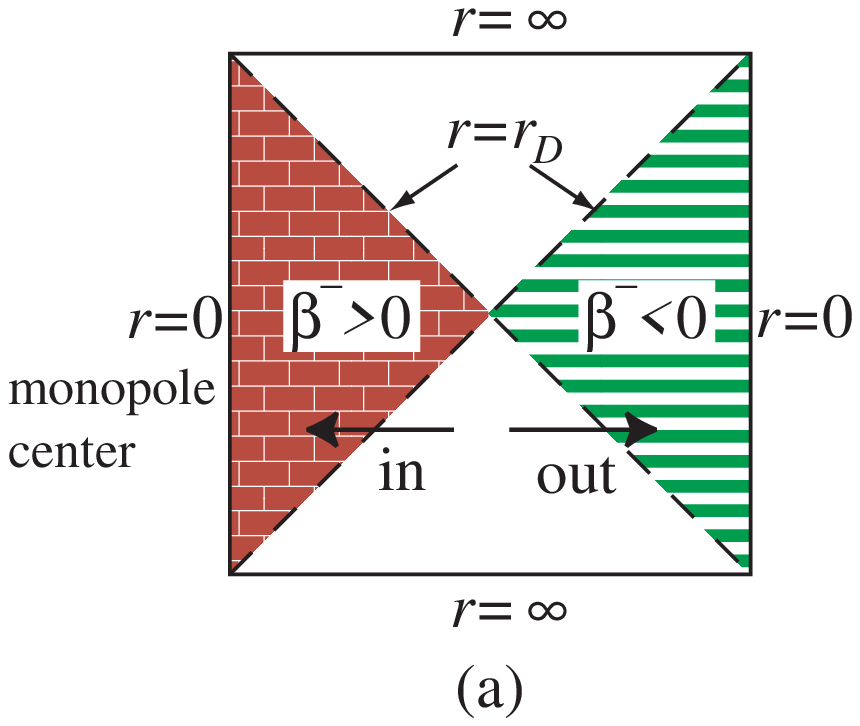}
  \psbox[scale=0.7]{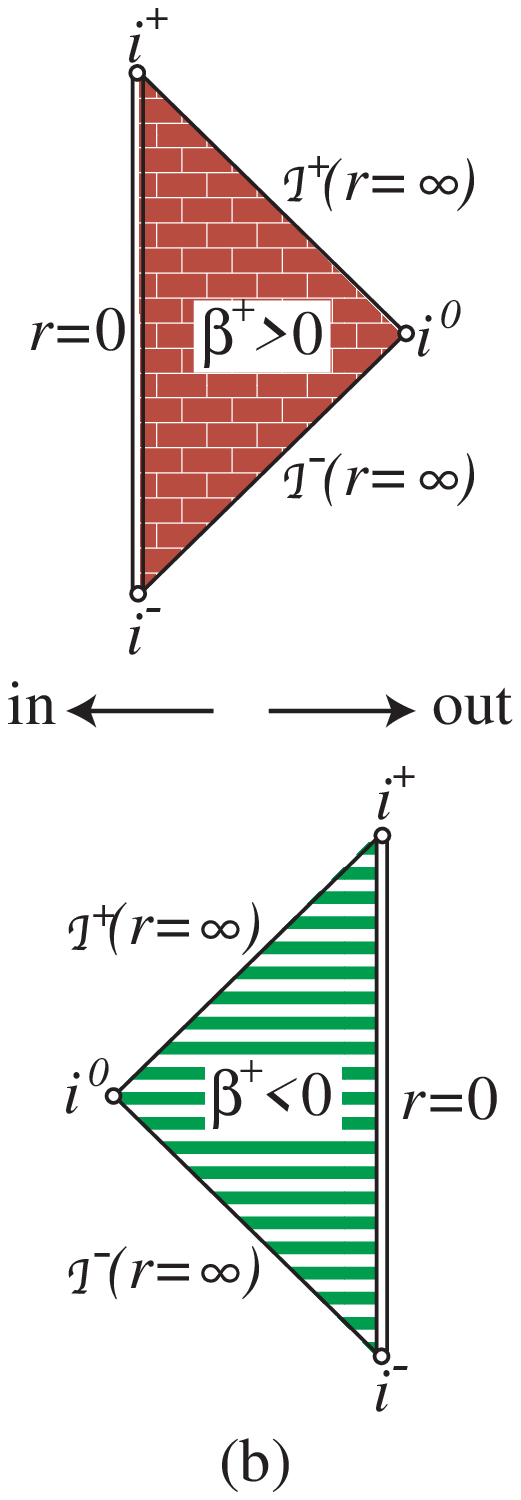}
  \psbox[scale=0.7]{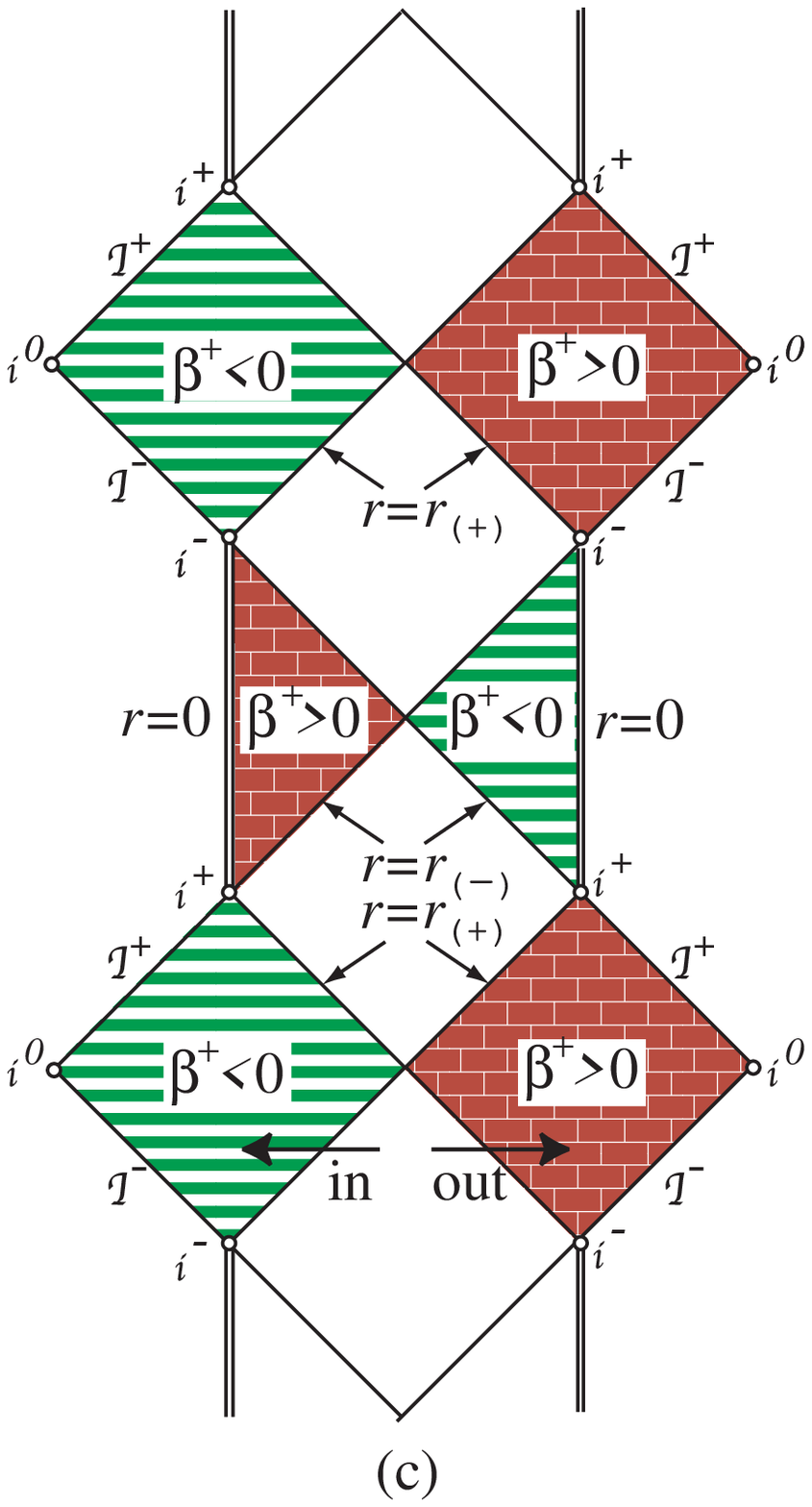}
 \end{center}
\end{figure}

\noindent
{\bf FIG.\ 1}. Conformal diagrams for (a) de Sitter spacetime, (b)
Reissner-Nordstr\"om spacetime with $m<q$, and (c)
Reissner-Nordstr\"om spacetime with $m>q$.  
$\cI^+$ and $\cI^-$ represent future and past null infinity, $i^+$ and 
$i^-$ represent future and past timelike infinity, and $i^0$
represents spacelike infinity. A double line denotes timelike 
singularity.
In Figs.\ 2-5 monopole solutions are described by joining a ``left" part 
of de Sitter spacetime to a ``right" part of Reissner-Nordstr\"om. 
The ``left" $r=0$ line in the diagram (a) corresponds to the monopole center.
Accordingly, the normal vector $n^{\mu}$ of $\Sigma$ points to the right in all diagrams. 
$r_D$ and $r_{(+)}~ (r_{(-)})$ denote de Sitter horizons and the black 
hole outer (inner) horizons, respectively.
In the region $r_-<r_D$ ($r_+>r_{(+)}$) the sign of 
$\beta^{-}=\pa_nr_{-}$ ($\beta^{+}=\pa_nr_{+}$) is definite
regardless of the shell motion.
Brick-pattern domains denote spacetime regions of positive-definite $\beta$, while 
stripe-pattern domains denote those of negative-definite $\beta$.

\begin{figure}
 \begin{center}
  \psbox[scale=0.53]{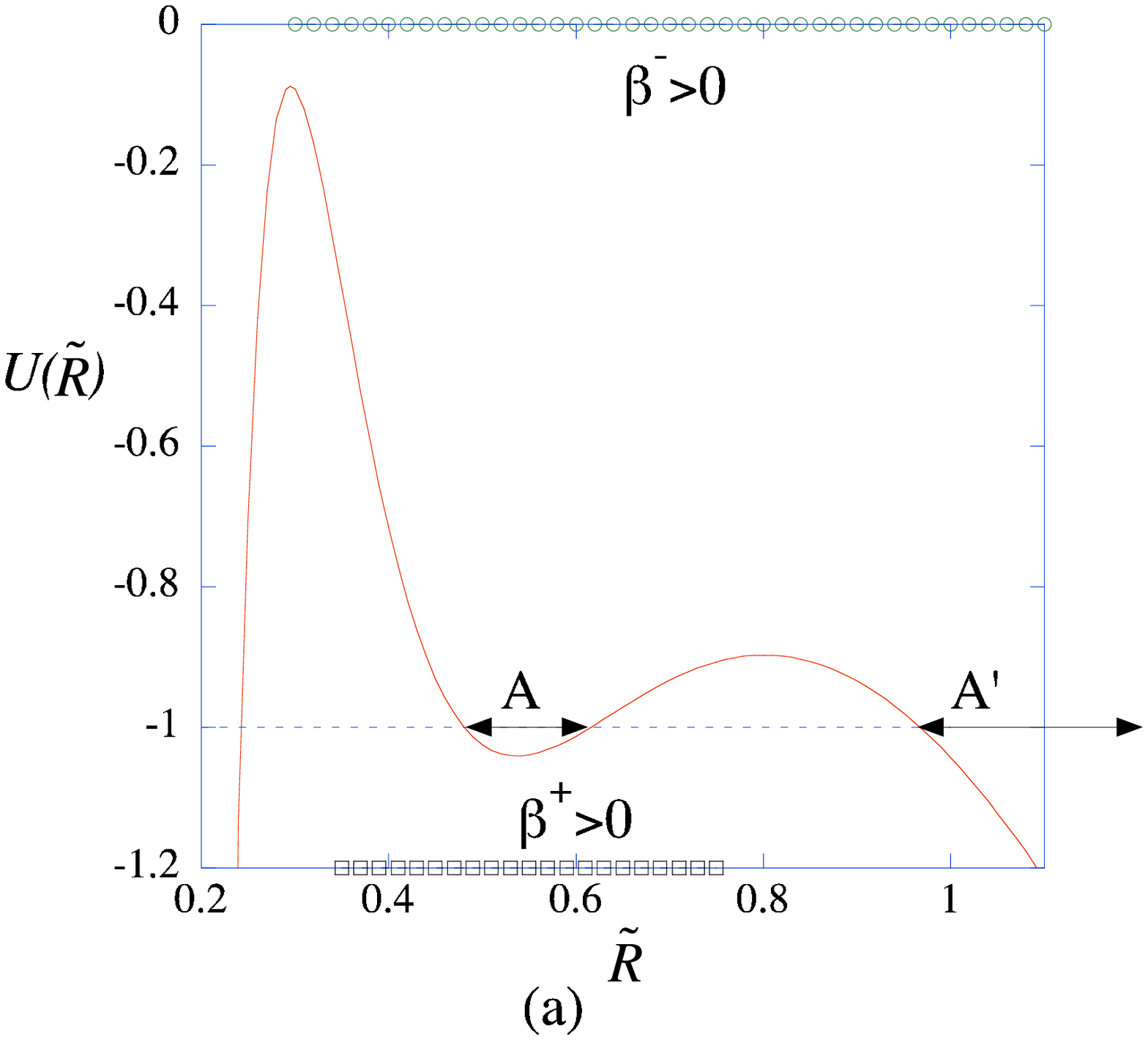}
  \psbox[scale=0.35]{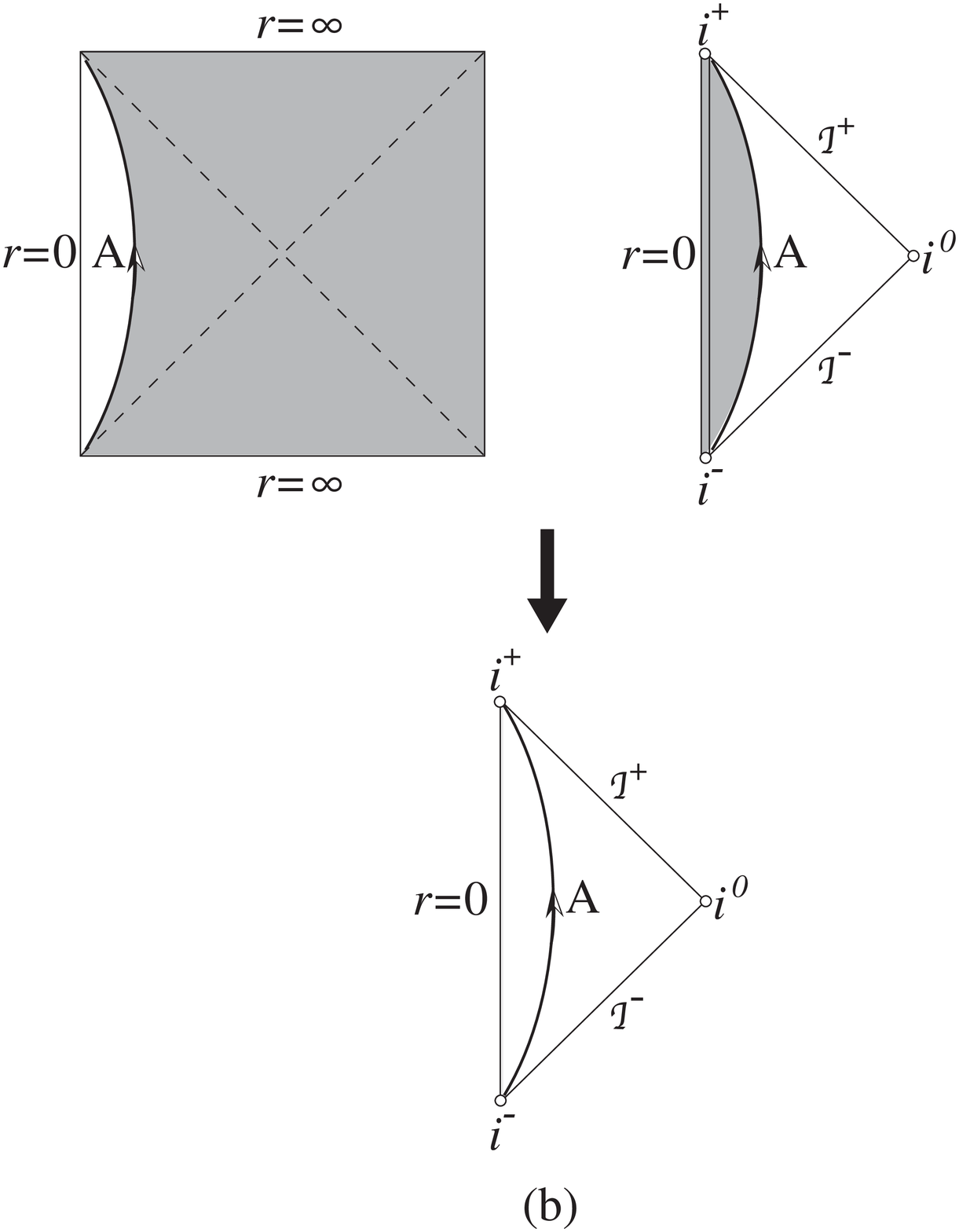}
 \end{center}
 \begin{center}
  \psbox[scale=0.35]{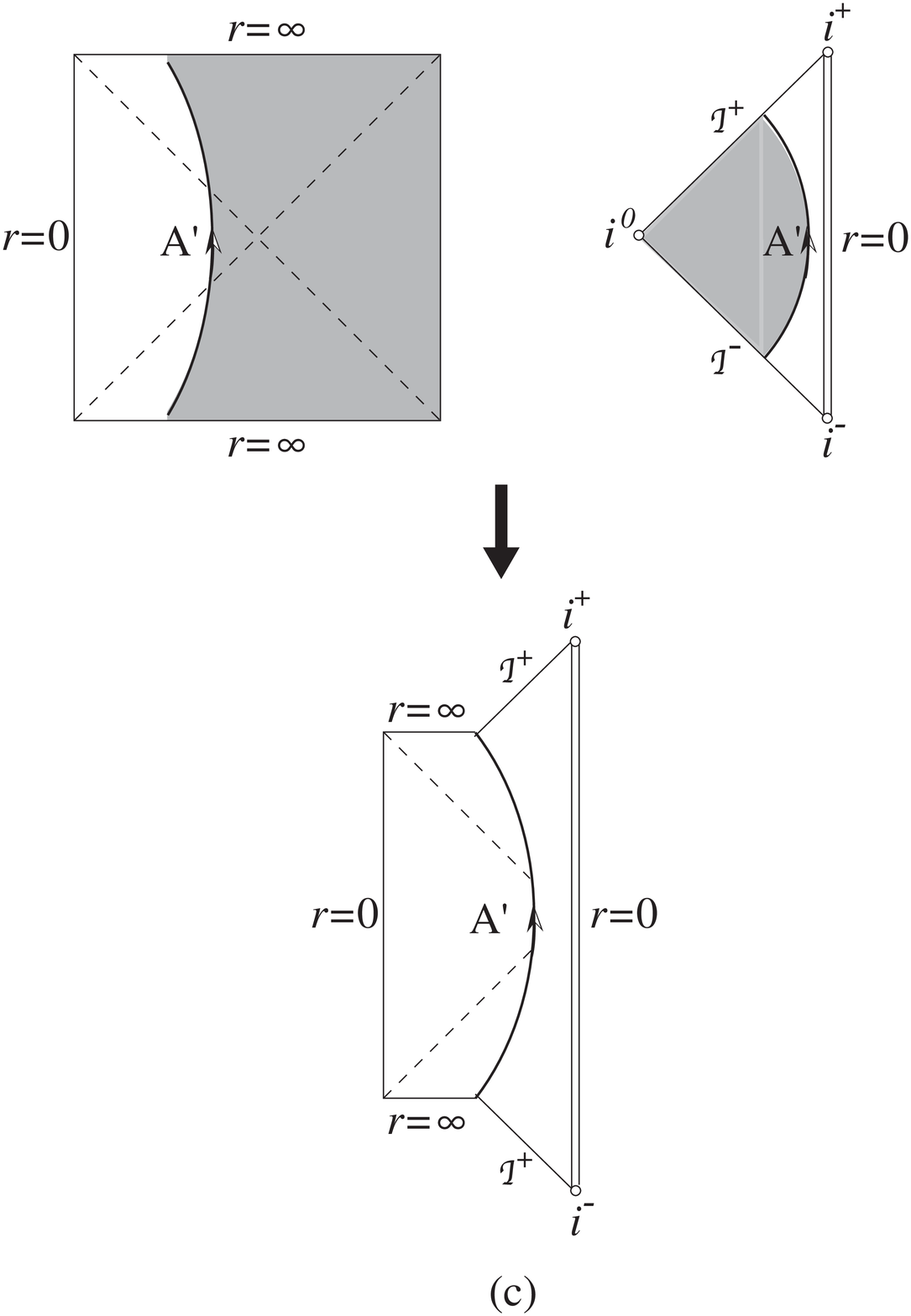}
 \end{center}
\end{figure}

\noindent
{\bf FIG.\ 2}. Solutions with $m=0.58,~q=0.6,~s_0=0.6$ and $s_1=0.1$. (a) 
represents the effective potential $U(\tr)$. There are a classically stable 
solution (type A) and a expanding solution (type A'). No black-hole horizon. 
Circles at the top denote the region of $\beta^->0$, while
squares at the bottom denote the region of $\beta^+>0$ .
(b) and (c) show the conformal diagrams of type A and A' solutions, respectively. 
The upper figures show how the trajectory of the shell is embedded in 
Reissner-Nordstr\"om full spacetime, where gray domains 
indicate nonexistent regions.
The lower figures show complete spacetimes.

\begin{figure}
 \begin{center}
  \psbox[scale=0.53]{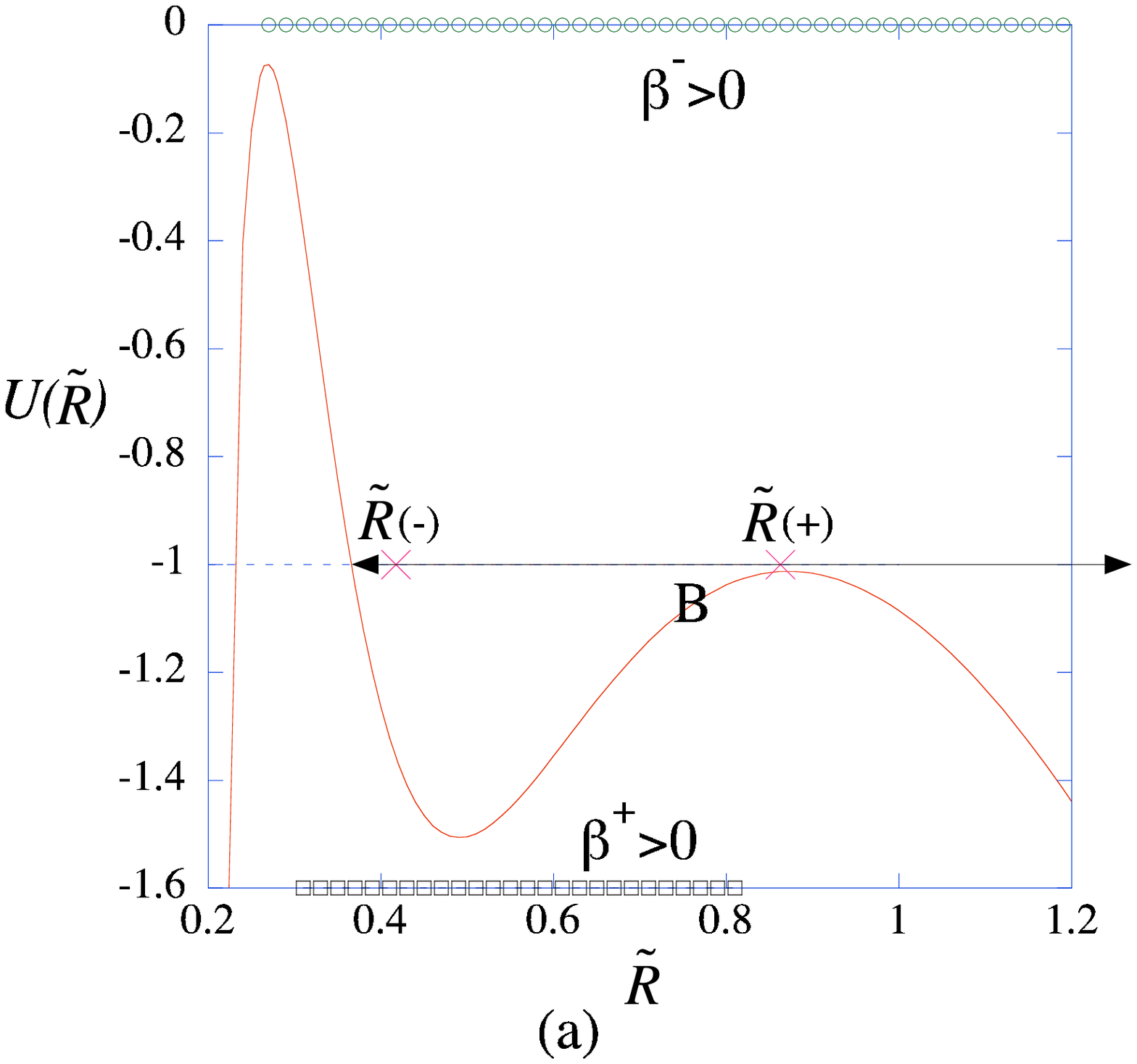}
 \end{center}
 \begin{center}
  \psbox[scale=0.33]{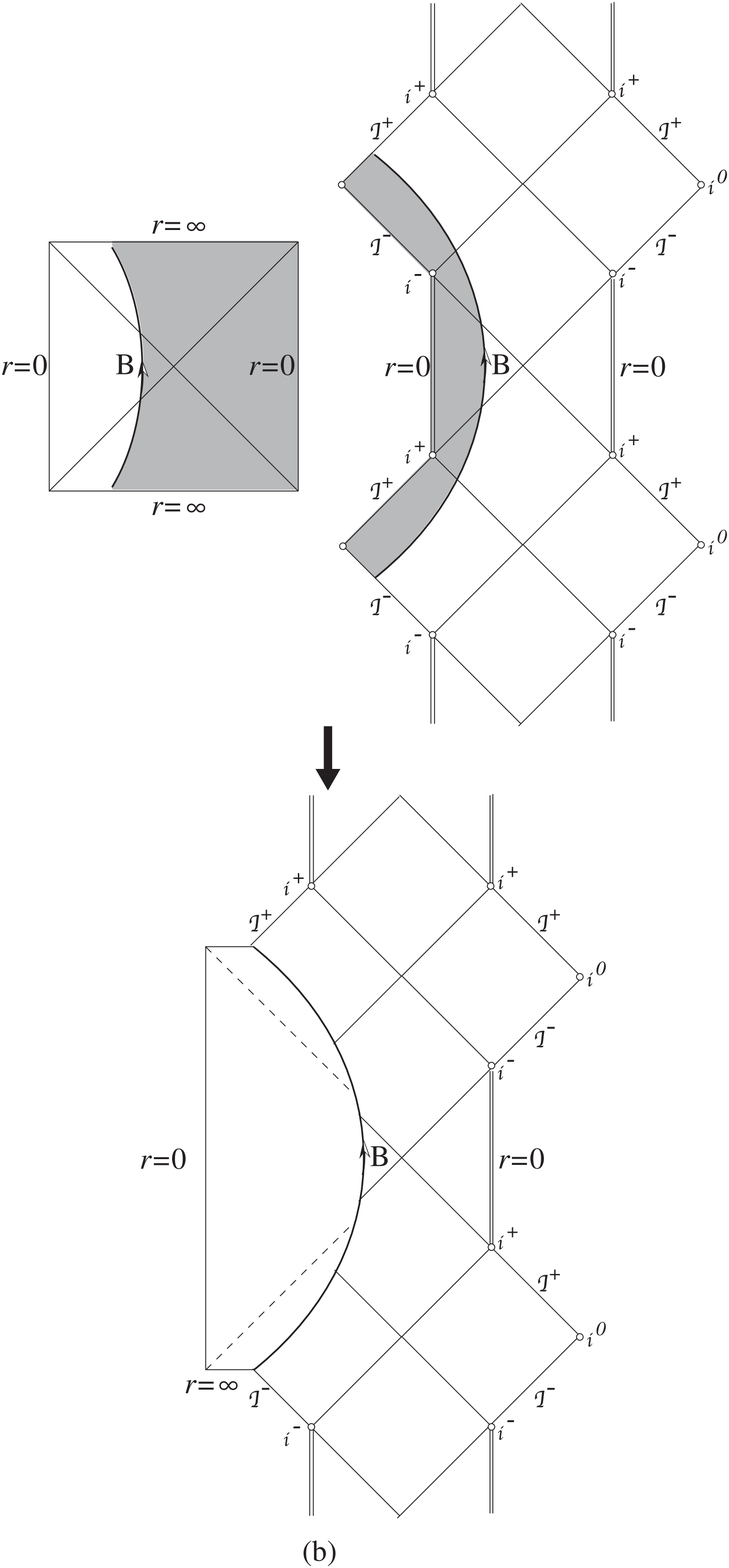}
 \end{center}
\end{figure}

\noindent
{\bf FIG.\ 3}. A solution with $m=0.64,~q=0.6,~s_0=0.6$ and $s_1=0.1$. (a) 
represents the effective potential $U(\tr)$. There is an inflationary
solution (type B). A cross denotes a black-hole horizon.
(b) shows the conformal diagram of the solution.

\begin{figure}
 \begin{center}
  \psbox[scale=0.33]{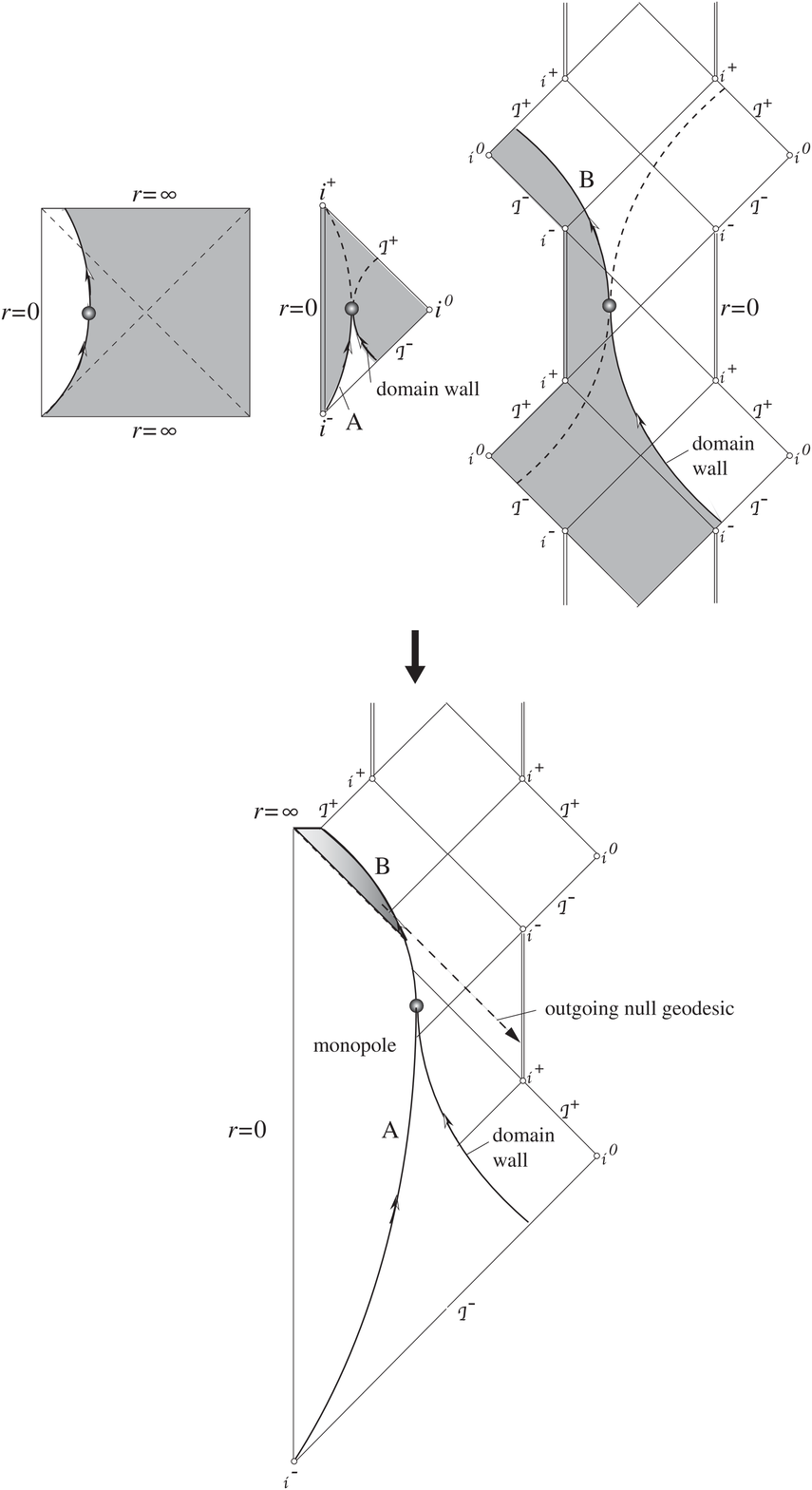}
 \end{center}
\end{figure}

\noindent
{\bf FIG.\ 4}. Conformal diagram of the model that a spherical domain 
wall surrounds the type A monopole and collides with it. A ball denotes 
an event of the collision. The stable monopole evolves into the type B 
inflating monopole by the collision.
The shaded domain denotes a set of anti-trapped surfaces in the de 
Sitter side, that is, the created inflationary universe.
The past directed outgoing radial null geodesics emanating from this 
region necessarily hit the timelike singularity.

\begin{figure}
 \begin{center}
  \psbox[scale=0.53]{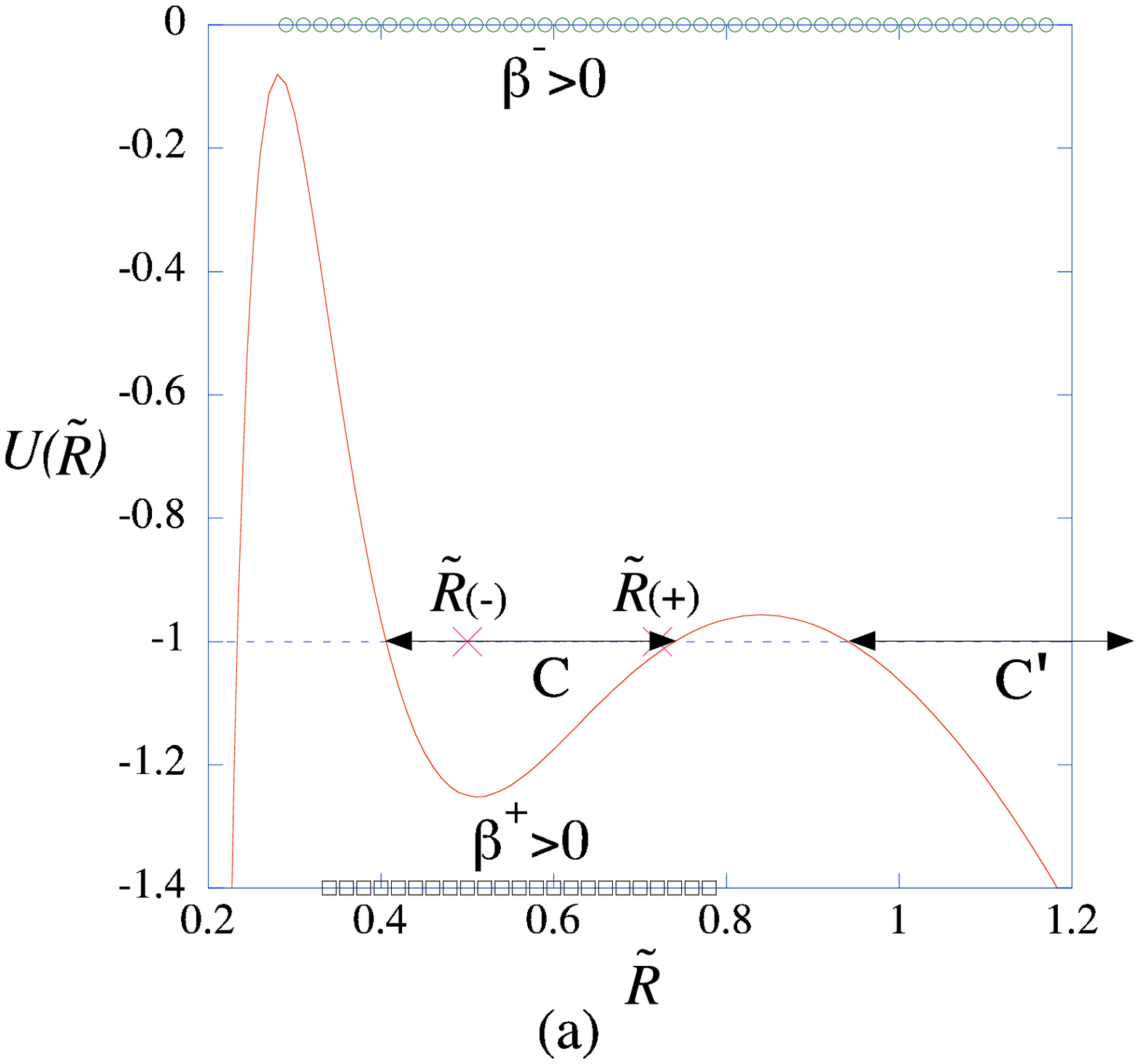}
  \psbox[scale=0.35]{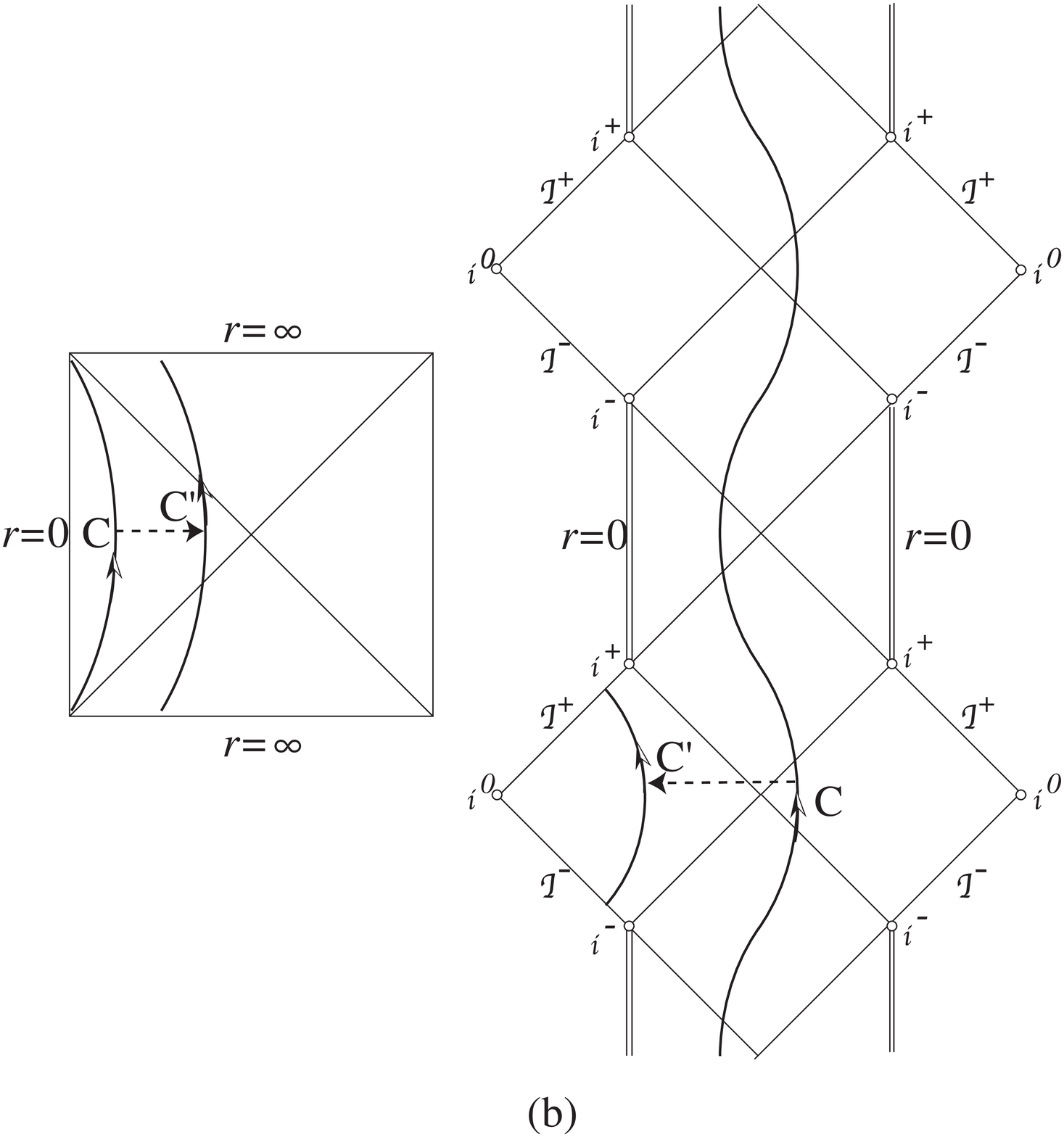}
 \end{center}
\end{figure}

\noindent
{\bf FIG.\ 5}. Solutions with $m=0.61,~q=0.6,~s_0=0.6$ and $s_1=0.1$. (a) 
represents the effective potential $U(\tr)$. There are a classically stable 
solution (type C) and an expanding solution (type C'). 
(b) shows the conformal diagram of the two solutions. C dashed line 
denotes a possible tunneling path.

\begin{figure}
 \begin{center}
  \psbox[scale=0.55]{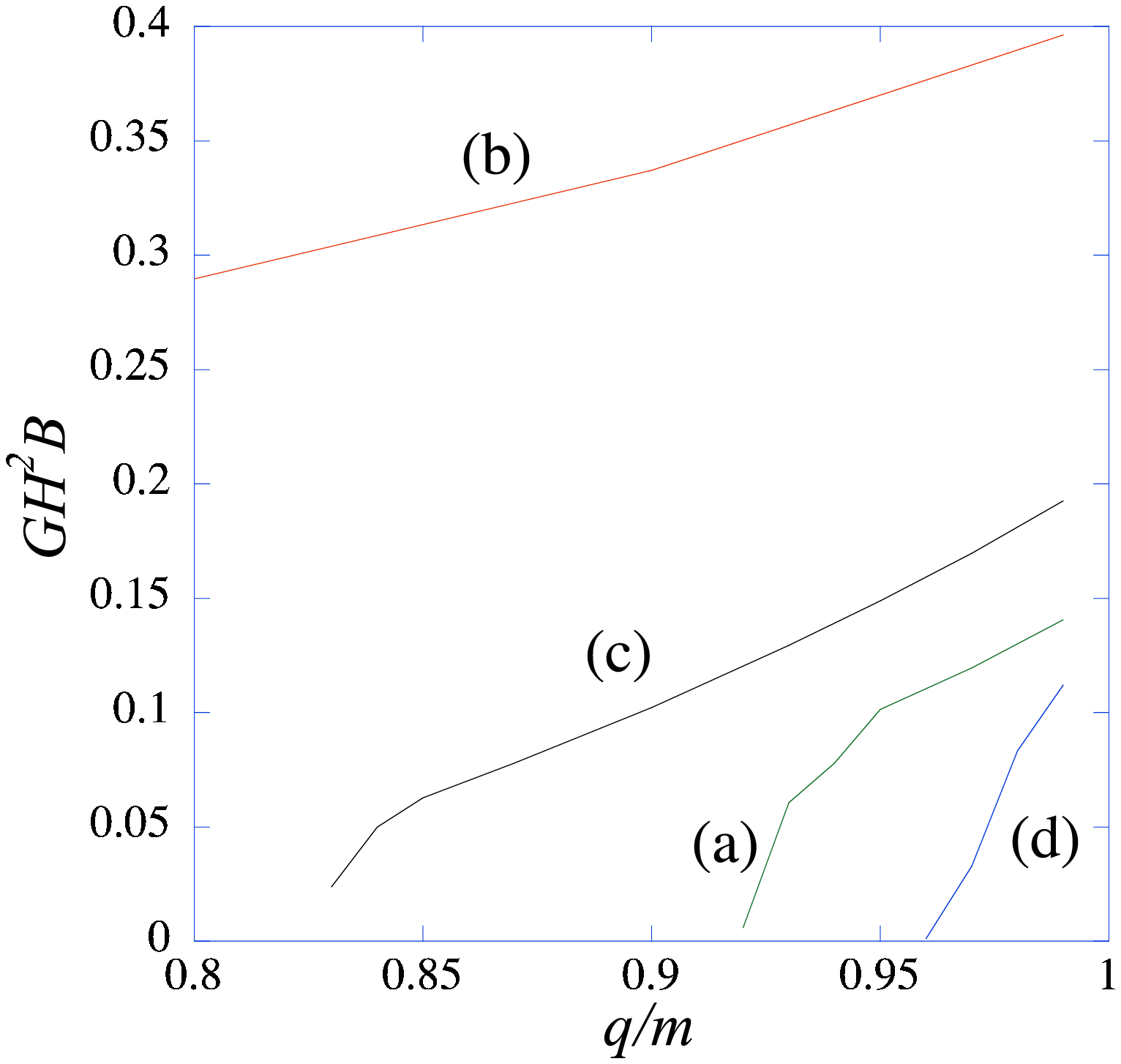}
 \end{center}
\end{figure}

\noindent
{\bf FIG.\ 6}. Prot of $GH^2B$ vs $q/m$ for several values of $m,~s_0$ and $s_1$.
(a) $m=0.6,~s_0=0.6$ and $s_1=0.1$;
(b) $m=0.4,~s_0=0.6$ and $s_1=0.1$;
(c) $m=0.6,~s_0=0.3$ and $s_1=0.1$;
(d) $m=0.6,~s_0=0.6$ and $s_1=0.2$.

\end{document}